\documentclass[%
 aip,
 reprint,
]{revtex4-2}

\usepackage{graphicx}% Include figure files
\usepackage{dcolumn}% Align table columns on decimal point
\usepackage{bm}% bold math

\usepackage[utf8]{inputenc}
\usepackage[T1]{fontenc}
\usepackage{mathptmx}
\usepackage{etoolbox}
\usepackage{amsmath,amssymb}
\usepackage{algorithm2e}

\usepackage[dvipsnames]{xcolor}

\SetKwComment{Comment}{$\triangleright$\ }{}

\makeatletter
\def\@email#1#2{%
 \endgroup
 \patchcmd{\titleblock@produce}
  {\frontmatter@RRAPformat}
  {\frontmatter@RRAPformat{\produce@RRAP{*#1\href{mailto:#2}{#2}}}\frontmatter@RRAPformat}
  {}{}
}%
\makeatother
\begin{document}

\preprint{AIP/123-QED}

\title[]{The Automated Bias Triangle Feature Extraction Framework}
% Force line breaks with \\
\author{M. Kotzagiannidis}
 \email[M. Kotzagiannidis: ]{madeleine.kotzagiannidis@mindfoundry.ai}
 \affiliation{Mind Foundry Ltd, Summertown, Oxford OX2 7DD, United Kingdom}%Lines break automatically or can be forced with \\
\author{J. Schuff}%
\affiliation{Department of Materials, University of Oxford, Oxford OX1 3PH, United Kingdom}
\author{N. Korda}%
 \affiliation{Mind Foundry Ltd, Summertown, Oxford OX2 7DD, United Kingdom}

\date{\today}% It is always \today, today,

\begin{abstract}
Bias triangles represent features in stability diagrams of Quantum Dot (QD) devices, whose occurrence and property analysis are crucial indicators for spin physics. Nevertheless, challenges associated with quality and availability of data as well as the subtlety of physical phenomena of interest have hindered an automatic and bespoke analysis framework, often still relying (in part) on human labelling and verification. 
We introduce a feature extraction framework for bias triangles, built from unsupervised,
segmentation-based computer vision methods, which facilitates the direct identification and quantification of physical properties of the former. Thereby, the need for human input or large training datasets to inform supervised learning approaches is circumvented, while additionally enabling the automation of pixelwise shape and feature labeling. 
In particular, we demonstrate that Pauli Spin Blockade (PSB) detection can be conducted effectively, efficiently and without any training data as a direct result of this approach. 
\end{abstract}

\maketitle

\section{\label{sec:level1}Introduction}

Quantum dot (QD) devices present a promising platform for scalable quantum computation \cite{scal1}\cite{art1}\cite{art2}\cite{art11}. Transport measurements in QD devices are performed by applying a bias voltage between source and drain reservoirs which draws a current through the device. 
Stability diagrams, which give information on the equilibrium number of charges in each dot, can be obtained via current measurements as a function of the plunger gate voltages $V_{P1}$ and $V_{P2}$ \cite{phys2}. \\ 
Bias triangles, occurring in pairs in stability diagrams, represent charge transitions and indicate that a device is tuned into the double quantum dot (DQD) regime. Visible stripes parallel to the common triangle base line characterize certain excited state transitions. These usually remain visible in both bias directions, and may differ in the magnitude of the corresponding current as well as in their position \cite{psb1}.\\
Many embedded physical properties of interest, such as excited state energy level spacings, lever arms, cross talk and detuning axes, can be extracted based on the knowledge and geometry of the triangles' (pixel) coordinates, including the base line, angles, and side lengths as well as prominent lines, enclosed segments thereof, and angle-bisector lines \cite{phys2}.
\\
Pauli Spin Blockade (PSB) is a crucial requirement for spin qubit initialisation and readout, providing a first stage indicator for potentially viable locations. Nevertheless, the visual detection of PSB presents a challenge even for experienced human labellers due to its subtlety. Concurrently, suitable data, to inform machine learning methods, is rare and imbalanced, with often only few available positive cases to learn from.
PSB is detected manually by comparing stability diagrams displaying bias triangles at different magnetic fields and inspecting changes in the current along the base line, manifested in current differences between the stability diagrams at this location. In particular, at zero magnetic field $B=0$, there is a blockade, visible in form of fainting or attenuation of current, at the base line of the bias triangles. For devices with strong spin-orbit interaction, this is lifted for a non-zero magnetic field $B\neq 0$ - other devices may exhibit the reverse effect \cite{phys3} \cite{phys2} \cite{psb1}. \\
To achieve the automated and quantitative analysis of physical properties exhibited in bias triangles, such as PSB, it is thus desirable to devise a local, pixel(-coordinate)wise approach in form of segmentation and feature extraction.
In the case of PSB, this enables a comparison of the pixel intensity within triangle segments enclosed by the base and excited state transitions for different magnetic fields.\\
\\
In this work, we establish a segmentation-based detection and feature extraction framework for bias triangles which facilitates the direct extraction of geometric elements and properties of the latter, ranging from base line detection to the occurrence of PSB, thereby automating tasks that have largely been conducted manually.\\
In summary, the proposed framework consists of the following steps
\begin{enumerate}
\item {\bf Shape approximation \& fitting} of basic detected contours to minimum-edge polygons for bias triangle segmentation. In particular, we propose 
\begin{enumerate} 
\item the relaxed Ramer-Douglas-Peucker (rRDP) algorithm, a variation of the classical RDP \cite{ramer}\cite{dp}, which specifically fits pairs of bias-triangles, and 
\item the Minimum Enclosing Triangle (MET) \cite{met1}\cite{met2} method, as an alternative for coping with device-specific bias triangles which lack the typical triangular structure, exhibiting noisy boundaries or disconnectedness. This approach a priori constructs a convex envelope of detected disconnected contours.
\end{enumerate}
\item {\bf Basic feature extraction}, including primarily the extraction of the bias triangle base line, excited state transitions, and individual segments (which can be extended to other relevant geometric shape properties).
\item {\bf PSB detection based on extracted features}: we make use of previously computed features to localize and compare intensity variations within triangle segments on pairs of bias triangles with blocked and unblocked current. Geometric properties extracted from the latter serve as the template and the classification is conducted with respect to a pixel intensity threshold.
\end{enumerate}
In prior work, the set of tasks revolving around the extraction of features in 2D scans, such as stability diagrams, has largely been tackled manually, and sometimes through the use of specialized signal processing methods, which are often rich in and sensitive to hyperparameters. More recently machine learning has been used, but these methods nevertheless rely on the availability of (large) training datasets
\cite{col1}\cite{art5}\cite{art6}\cite{art7}.\\
A neural-network-based approach\cite{psb1} for the classification of PSB occurrence in bias triangles employing cross-device validation was developed; nevertheless, classification is conducted on the image-level, thereby not directly facilitating a local, pixel-wise approach for further feature extraction. Moreover, an auto-encoder\cite{bt2} was employed to score the quality of bias triangles with respect to changing voltage configurations.
The bias triangle-module of the Quantum Technology Toolbox (QTT)\cite{qtt} facilitates the extraction of few properties such as lever arms, however, it is primarily a drawing tool which requires user-input in form of the delineation of boundary points, with a limited range of extractable features.\\
In contrast, here, we consider an unsupervised approach which employs computational geometry with a small set of hyperparameters that require minimal adjustment. To the best of our knowledge, there is no comparable method which facilitates the full range of automatic feature extraction based on segmentation for bias triangle analysis.\\
\\
We exemplify the proposed approaches on both simulated and experimental data, respectively stemming from a physics-based simulator and different FinFET devices, which constitute publicly available datasets \cite{psb1}\cite{data}.
Ultimately, we demonstrate high PSB classification scores, with accuracies ranging from mid-to high-80\%s across the various datasets and devices, with relatively low numbers in false negative and false positive detections. In addition, we report robust segmentation accuracy, exemplified through segmentation scores, such as a Dice coefficient of 87.03\% on a simulated dataset with 500 randomly generated examples, which encompasses a wide variety of triangle shapes, as well as through a selection of visuals for experimental datasets of diverse quality and characteristics.\\
\\
The paper is organized as follows: we begin by outlining the proposed methods in Sect.\ II, followed by results and discussion on both simulated and experimental datasets of bias triangles in Sect.\ III, and provide concluding remarks and an outlook in Sect.\ IV. 

\section{Methods}
In the proposed bias triangle framework, we can automatically extract properties such as base lines, angles and segments enclosed by prominent (excited state) lines, by employing computational geometry methods which approximate detected contours with minimum-edge polygons.
Furthermore, the extraction of such local properties is built upon to function as a classification tool for physical phenomena such as PSB.

\subsection{Segmentation}
\begin{figure}
\vspace{1cm}\hspace{-7.99cm}
\includegraphics[scale = 0.52]{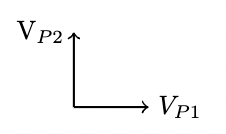}\\
\vspace{-2.75cm}%2.9
\includegraphics[scale = 0.32]{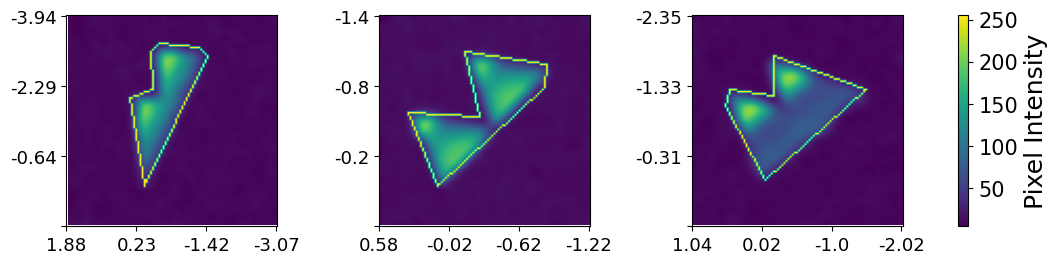}% 
\caption{\label{fig:first} Processed images of different simulated bias triangles (overlaid in plunger gate voltage space of $V_{P1}$ and $V_{P2}$, in V), which have been automatically segmented using the proposed rRDP method.}
\end{figure}

\begin{figure}
\includegraphics[scale=0.325]{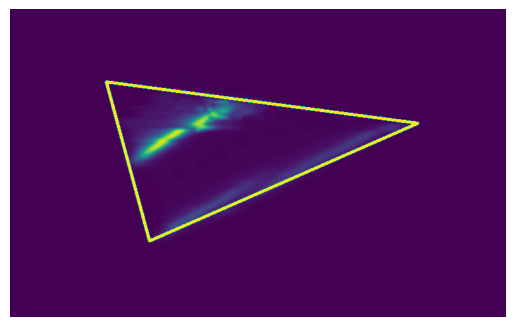}% 
\includegraphics[scale=0.3]{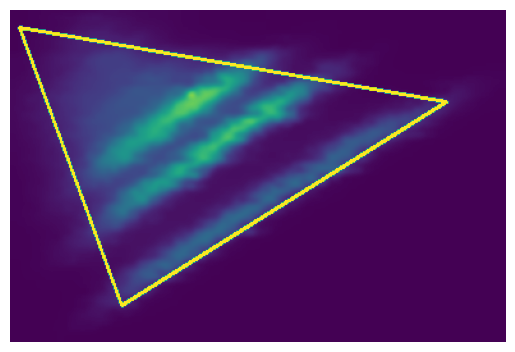}
\caption{\label{fig:third} Segmented bias triangles obtained via the MET method from the FinFET dataset (same device).}
\end{figure}

Given a stability diagram scan of current, the proposed segmentation framework performs binary thresholding on the image, followed by an initial detection of contours to which a polygon approximation routine is applied. In particular, by employing a relaxed version of the Ramer-Douglas-Peucker (RDP) algorithm, the detected shapes (bias triangle pairs) are fitted and refined to describe their prominent edges. Fig.~\ref{fig:first} depicts some examples of bias triangles and their segmentations, whereby axes have been universally converted from plunger gate voltage space to pixels (with current normalized and converted to pixel intensity, ranging from 0 to 255); for consistency, we depict all triangles as pointing to the top left corner.
%Fig.~\ref{fig:first}%
Under the assumption that the bias triangle scan is of low resolution (i.e.\ less than 100 pixels along any axis), it is preprocessed prior to thresholding (using, among other methods, Gaussian filtering) and interpolated by a specified resolution factor, in order to ensure good segmentation performance.
The thresholding function may be varied as per device-data characteristics (if known) - the triangle-method\cite{tr} is particularly well-suited for images whose histograms feature a dominant background peak, while the mean method, which uses the mean value of all pixels to determine the threshold value, may be suited for instances with some illumination (within or outside of the shape, creating noisy boundaries).\\
\\
The RDP is a discrete curve approximation algorithm, which decimates the number of line segments connecting points on it according to a specified precision, such as a fraction $\epsilon$ of the total length or perimeter\cite{ramer}\cite{dp}. As it can be employed for polygon approximation with the objective to minimize the number of edges, we utilize this technique as a classification metric, in particular, with the number of edges of a pair of bias triangles functioning as an unambiguous identifier. In contrast to other segmentation algorithms, the reduction (and counting) of edges presents an additional feature which not only provides a robust shape identifier but also enables fitting of imperfect triangles to facilitate property extraction. We proceed to describe the methods in more detail.
\\
{\bf The relaxed RDP (rRDP).}
We relax the RDP by setting an initial approximation tolerance $\epsilon$, as a fraction of the total perimeter, for the shape at hand, which is gradually increased until we reach either of two stopping criteria: if the Intersection over Union (IoU) score between original detection and approximation falls below a certain value (85\%) or the number of edges falls below 6 - the latter is motivated by the fact that bias triangles occur in (overlapping) pairs of two and would strictly possess a minimum of 5 edges. \\
If the detected shape consists of disconnected components, we allow the option to employ an alternative shape approximation algorithm, the Minimum Enclosing Triangle (MET) method\cite{met1}\cite{met2}, which considers the convex area enclosing the detected components. This phenomenon of disconnectedness is exhibited here in the FinFET data which features prominent lines within a triangle, yet with vanishing boundaries and interior intensity.
Different examples for this approximation are shown in Fig.~\ref{fig:third}.
\noindent In order to prevent confusion between disconnected components that are noise-related outliers or belong to separate bias triangles, and those that belong to the same pair, when this option is enabled, we further remove outliers at the image boundaries as well as enforce comparisons of the total area and areas of intersection (via IoU) between different approximations, to gauge whether the correct shape is approximated. A basic version of the method is outlined in pseudocode in Algorithm \ref{alg1}.

\RestyleAlgo{boxruled}
\LinesNumbered
\begin{algorithm}[ht]
  \caption{Triangle Segmentation Algorithm\label{alg1}}
 {\bf Inputs:} image $I$, {\bf parameters:} $R$, $area_{\mathrm{min}}$, $area_{\mathrm{max}}$, \\
 $\mathrm{Thr}_{\mathrm{method}}()$, $OPT_{\mathrm{MET}}$\\
 Image pre-processing \& interpolation by resolution factor $R$: $I_{\mathrm{R}}$\\
Binary thresholding: $I_{\mathrm{T}}$ = $\mathrm{Thr}_{\mathrm{method}}$ ($I_{\mathrm{R}}$)\\
Detect contours: $C = \{C_i\}_i$ = sort (contour($I_{\mathrm{T}}$))\\
 \uIf{$\mathrm{area}(C_i)> area_{\mathrm{min}}\enskip \&\enskip \mathrm{area}(C_i) < area_{\mathrm{max}}$}{\Comment*[r]{Remove outliers}
 \uIf{$\mathrm{components}(C)>1\enskip \&\enskip OPT_{\mathrm{MET}} = \mathrm{True}$ }{
Apply {\bf MET}:\\
        Get extreme points of \textit{C}: $\tilde{C} = \mathrm{corners}(C)$\\
        $\textit{C}_{\mathrm{approx}} = \mathrm{MET}(\tilde{C})$\\
}
  \uElse{
  Apply relaxed {\bf RDP}:\\ 
  \While{$\textit{no}_{\mathrm{edges}}> 6 \enskip \&\enskip overlap >0.85$}{
           $\textit{C}_{\mathrm{approx}} = \mathrm{RDP}(C,\epsilon *\mathrm{perimeter}(C))$\\
            $\textit{no}_{\mathrm{edges}} = \mathrm{len}(\textit{C}_{\mathrm{approx}})$\\
             $overlap = \mathrm{IoU}(\textit{C},\textit{C}_{\mathrm{approx}})$\\
             $\epsilon = \epsilon + 0.001$\\
             } 
  }
  }

{\bf Output}: Segmentation $C_{\mathrm{approx}}$   \\ 
\end{algorithm}

\subsection{Feature Extraction \& PSB-detection}
\begin{figure*}
\includegraphics[scale = 0.29]{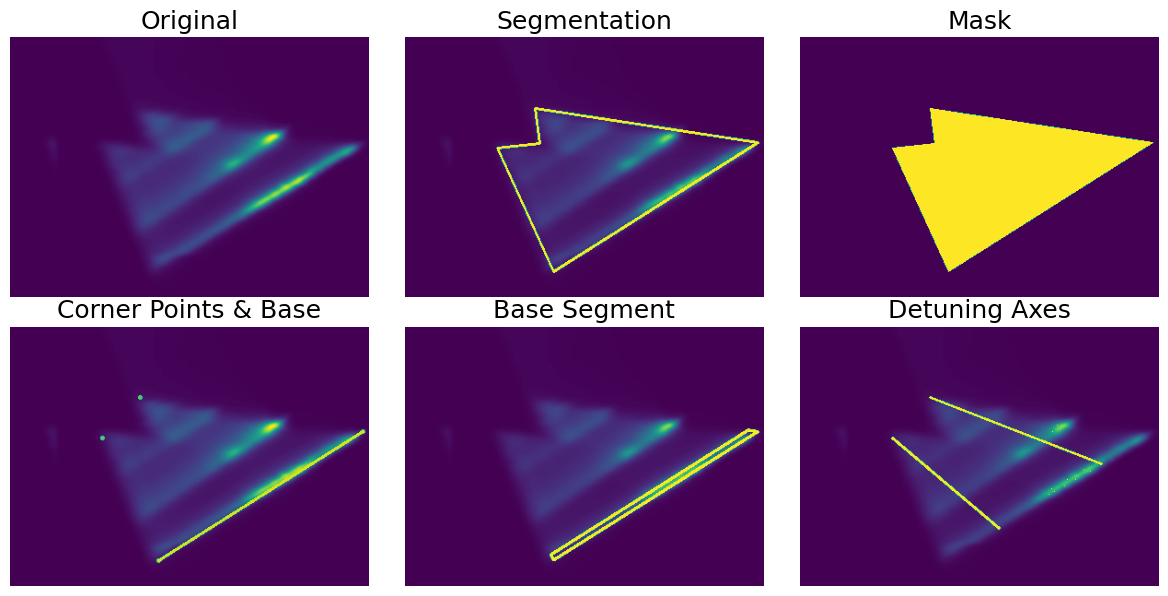}\hspace{5mm}
\includegraphics[scale = 0.29]{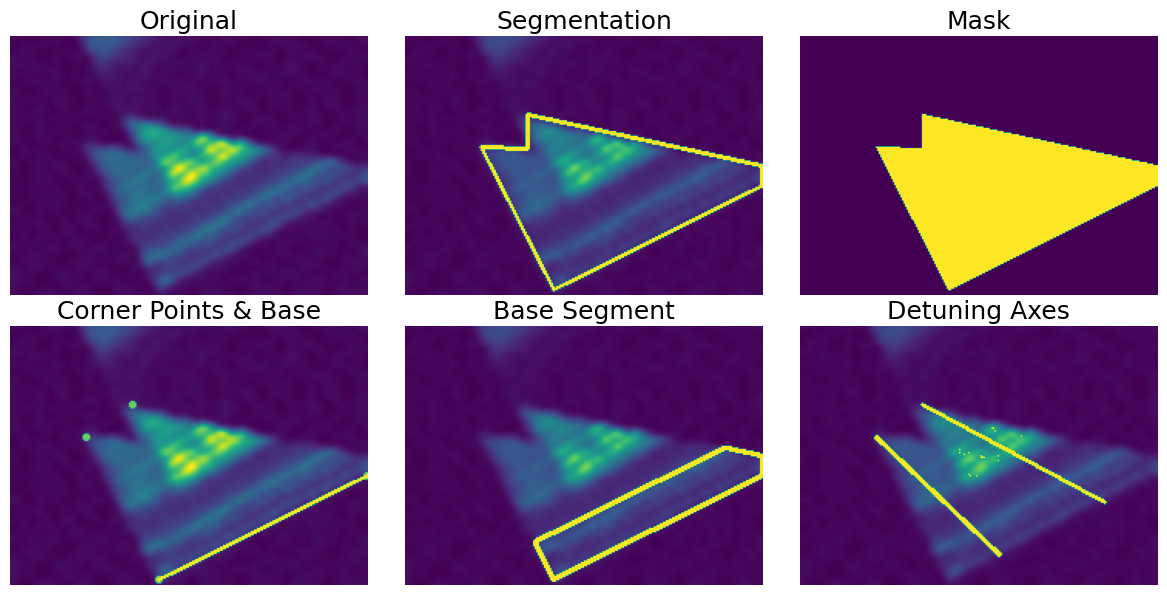}
\includegraphics[scale = 0.29]{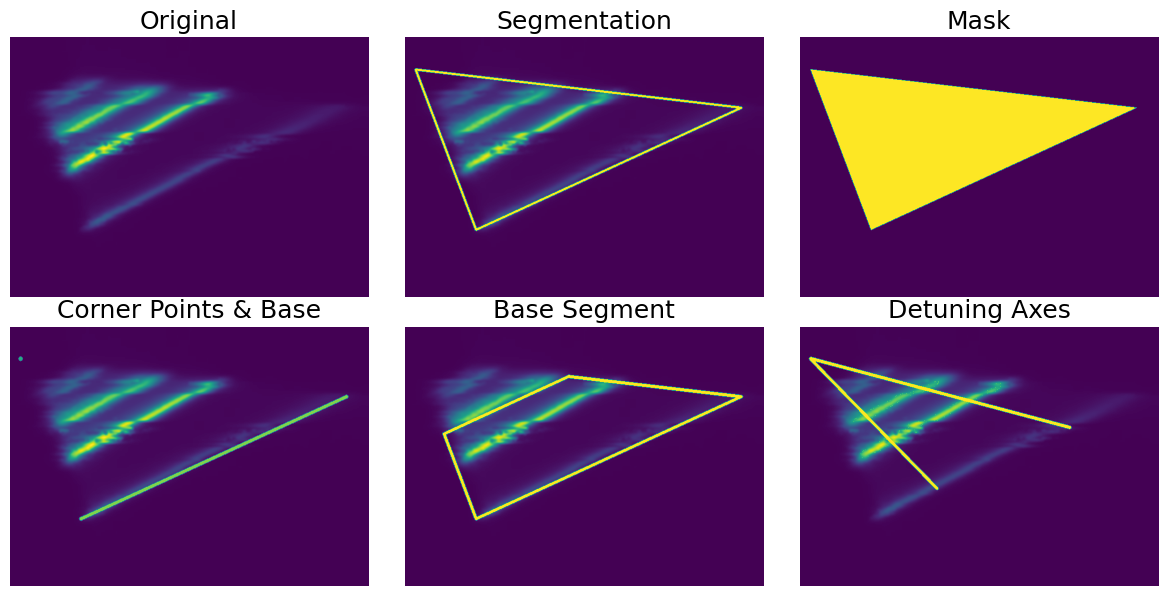}\hspace{7mm}
\includegraphics[scale = 0.299]{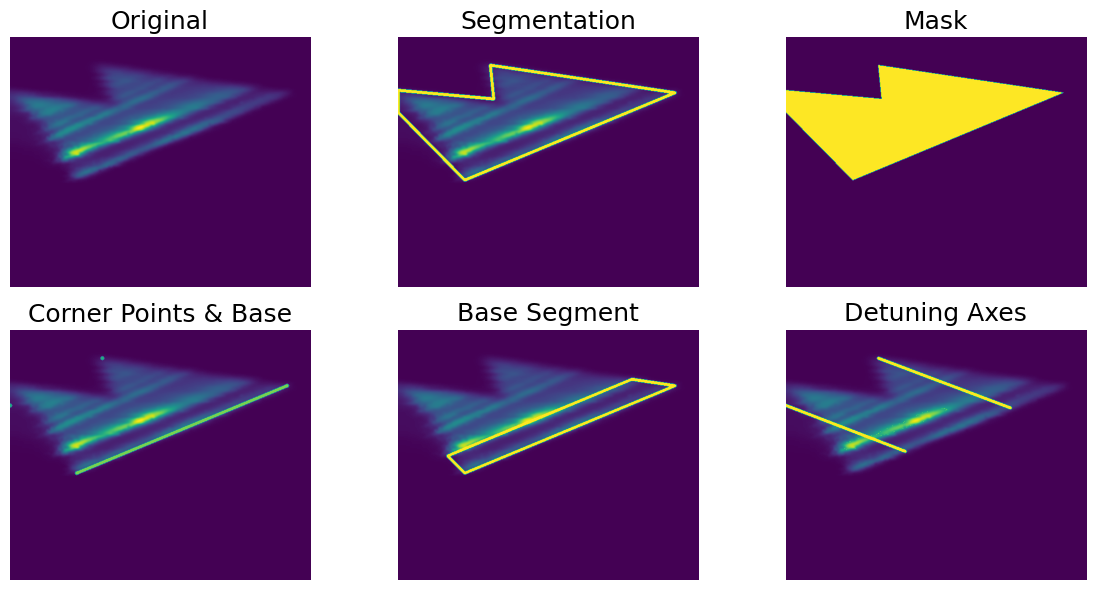}
\caption{\label{fig:wide}Examples of segmented bias triangles with extracted geometric features (FinFET). We respectively depict the original segmentation outline, pixelwise mask, followed by the extracted corner points and base line, base segment and detuning axes.}
\end{figure*}

\begin{figure*}
\includegraphics[scale = 0.3]{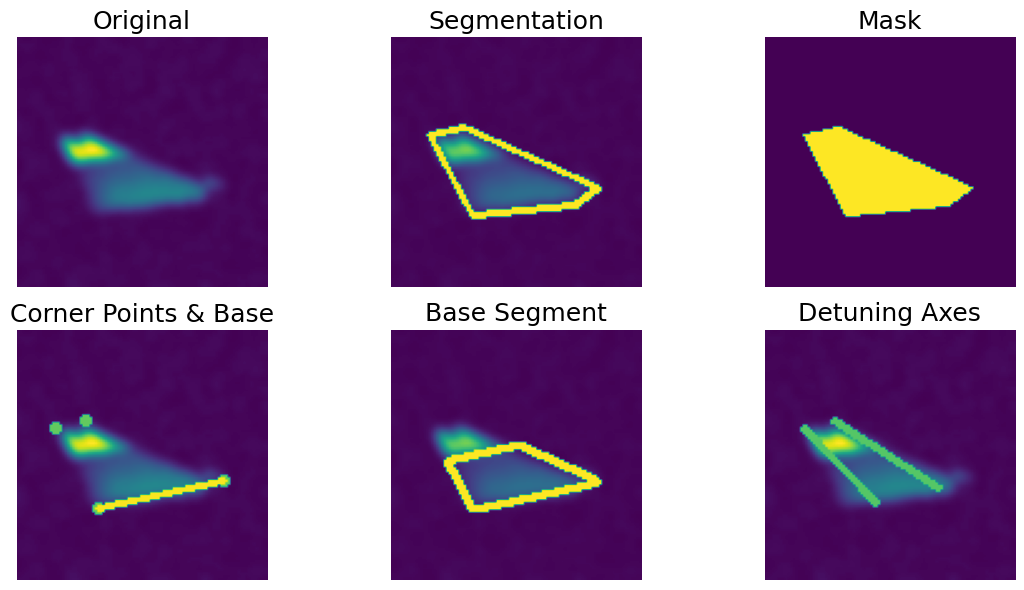} \hspace{8mm}
\includegraphics[scale = 0.3]{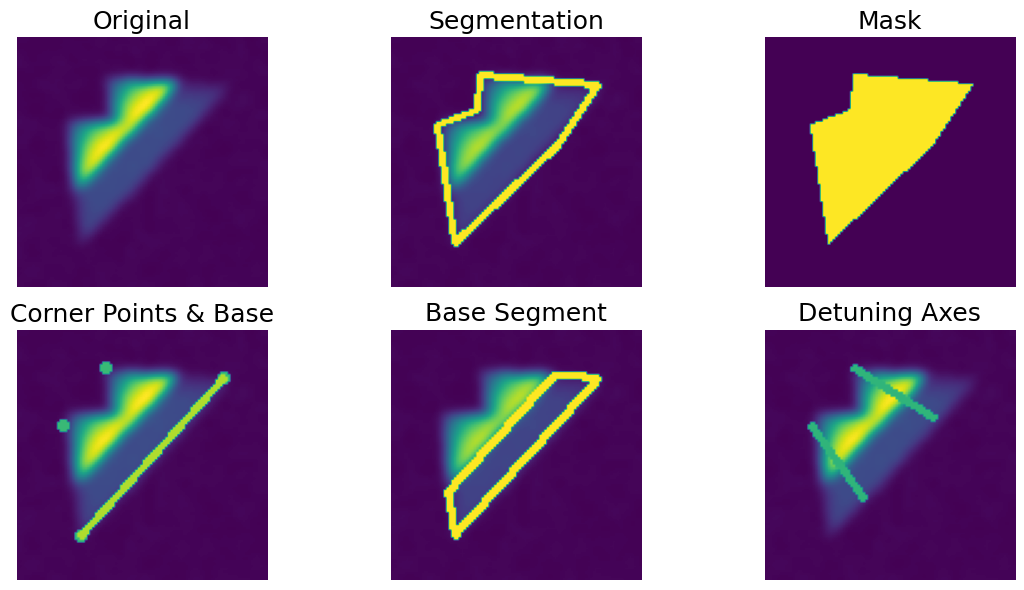}
\includegraphics[scale = 0.3]{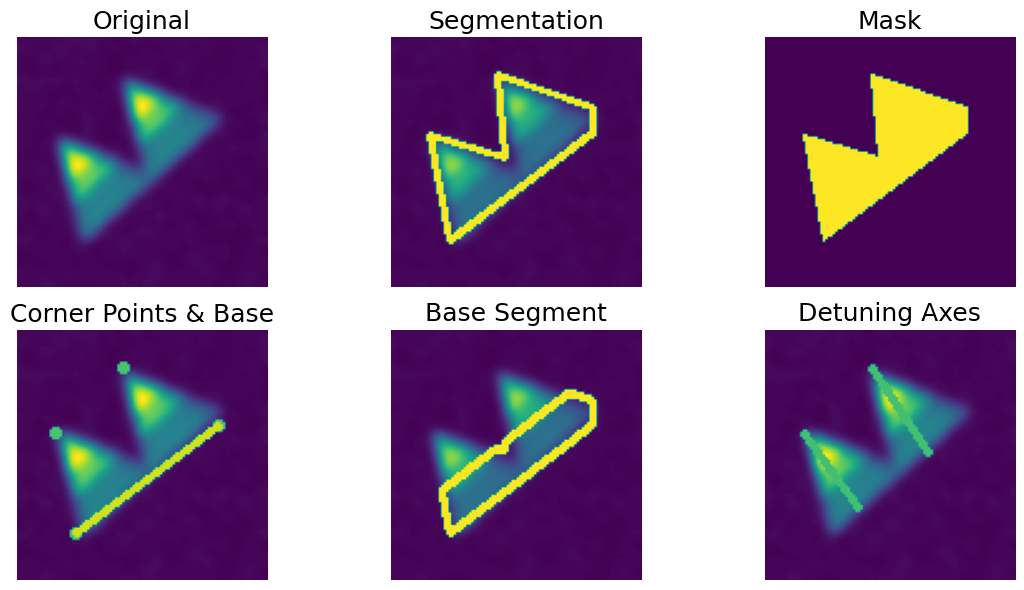}\hspace{8mm}
\includegraphics[scale = 0.3]{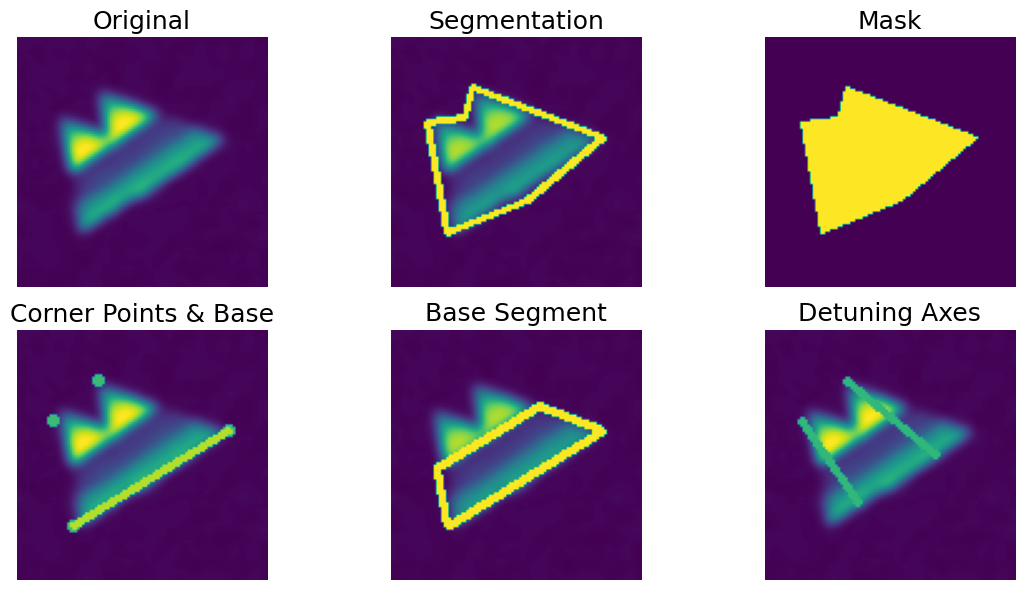}
\caption{\label{fig:wide2}Examples of segmented bias triangles with extracted geometric features (Simulated). We respectively depict the original segmentation outline, pixelwise mask, followed by the extracted corner points and base line, base segment and detuning axes.}
\end{figure*}

Following the polygon fitting to obtain a pixelwise localization of the bias triangle pairs, we proceed to extract properties such as the coordinates of the base line (as the orientation of the triangle is known to the experimenter in advance) and characteristic lines within the triangle. 
Some examples for a selection of extracted features for different datasets are depicted in Figs.~\ref{fig:wide} and ~\ref{fig:wide2}.
%Fig.~\ref{fig:wide}%

The precise knowledge of the edges and base further facilitates the computation of slopes and other interior triangle distances (corresponding to cross-talk and lever arms\cite{phys2}) as well as other properties pertaining to distances between excited state lines. For brevity, we summarize two feature extraction steps in Algorithm \ref{alg2}; other geometric properties can be similarly extracted on this basis.

\RestyleAlgo{boxruled}
\LinesNumbered
\begin{algorithm}[ht]
  \caption{Feature Extraction: Base \& Excited State Lines \label{alg2}}
{\bf Inputs:} image $I$, $triangle\_direction$, segmented shape $C$\\
Get extreme points $\tilde{C} = \mathrm{corners}(C)$\\
Compute distances $\{D_i\}_i$ along polygon between $\tilde{C}\setminus{D}$, where $D\in\tilde{C}$ aligns with the $triangle\_direction$.\\
Select pair with $D_{\mathrm{max}}$ (or by sign of slope, if known)\\
{\bf Outputs}: pair of points forming base line $l_{\mathrm{B}}$\\
 \hspace{0.1mm}
\hrulefill\\
 
{\bf Inputs:} image $I$, $triangle\_direction$, $slope\_tol$, $l_{\mathrm{B}}$, $Median\_OPT$ \\
Find all lines $l_i$ in scan: $\{l_i\}_i$ = FLD($I$)\\
\uIf{\emph{abs}(\emph{slope}($l_i$)-\emph{slope}($l_{\mathrm{B}}$))< slope\_tol}{Fit $l_{\mathrm{i}}$ parallel to base: $\tilde{l_{\mathrm{i}}}$}
\uElse{Discard $l_{\mathrm{i}}$}
\uIf{triangle\_direction = 'up' or 'right'}{
Keep $\tilde{l_{\mathrm{i}}}$ with $y_{intercept}(l_{\mathrm{B}}) < y_{intercept}(\tilde{l_{\mathrm{i}}})$}
\Else{Keep $\tilde{l_{\mathrm{i}}}$ with $y_{intercept}(l_{\mathrm{B}}) > y_{intercept}(\tilde{l_{\mathrm{i}}})$}
If $Median\_OPT$ = True, select median of sort($y_{intercept}(\tilde{l_i})$), else select the maximum\\
{\bf Outputs}: pair of points forming excited state line $l_{\mathrm{E}}$
\end{algorithm}
\noindent We note that the identification of the excited state lines can be further refined if e.g.\ specific transitions of interest and/or at a certain distance from the base need to be extracted; here, we defer to a data-driven heuristic in form of the median or outmost detected line in order to capture the intensity distribution of triangle segments automatically.

\RestyleAlgo{boxruled}
\LinesNumbered
\begin{algorithm}[ht]
  \caption{PSB detection algorithm\label{alg3}}
{\bf Input}: Image $I$, triangle contour $C_{\mathrm{approx}}$ from unblocked scan, base $l_\mathrm{B}$, excited state line $l_\mathrm{E}$, $int\_tol$\\
\uIf{Data \emph{is} FinFET \emph{or} Simulated}{
Select $Median\_OPT$ = True in Algorithm \ref{alg2}\\
}
\uElseIf{Data \emph{is} Other}{
Select $Median\_OPT$ = False in Algorithm \ref{alg2}\\
}

Extract triangle segment $S$ enclosing $l_\mathrm{B}$ and $l_\mathrm{E}$\\
Compute mean pixel intensities  $I_\mathrm{U} = \mathrm{mean}(S)/\mathrm{mean}(C_{\mathrm{approx}})$ over unblocked scan\\
Compute mean pixel intensities  $I_{\mathrm{B}} = \mathrm{mean}(S)/\mathrm{mean}(C_{\mathrm{approx}})$ over blocked image\\
{\bf Optional}: if dist($l_{\mathrm{B}}$, $l_{\mathrm{E}}$)< $seg\_tol$ * size($I$), $PSB$ = False\\
\uIf{$(I_{\mathrm{U}}-I_{\mathrm{B}})>int\_tol$}{
     $PSB$ = True}
\uElse{
     $PSB$ = False}
%\uEndIf
{\bf Output}: prediction $PSB$, intensity pair $(I_{\mathrm{U}}, I_{\mathrm{B}})$
\end{algorithm}
\noindent {\bf The PSB-detection framework.} We design the unsupervised PSB-detector as follows: given a bias-triangle 2D scan exhibiting unblocked current (and hence prominent base line), we apply the segmentation framework and extract its fitted polygon, along with the segment enclosing the base and prominent excited state line.
Here, potential lines are initially detected through OpenCVs Fast Line detector (FLD) \cite{fld14}, and further fitted to be parallel to the base line. For the best choice among detected (excited state) lines, we employ a heuristic that is device-data-specific and ensures additional robustness to faulty line detection: depending on the bias, bias triangles may contain multiple detectable lines, so in order to obtain a segment that is not too small or large, we select the median of the detected lines (as ordered by their y-intercept). 
On the other hand, bias triangle data taken at different temperatures and/or from a different device may feature fewer, less spread out lines at smaller distances to the base, so in order to ensure a large enough segment, we choose the outmost detected line in such a case. 
Subsequently, the algorithm superimposes the detected contours on the scan exhibiting blocked current and proceeds to respectively extract the average pixel intensity of the segment normalized by the total triangle intensity for both images. The normalization is conducted to account for (uniform) pixel intensity variations within the triangle between scans as well as to obtain relatively comparable values. Following the specification of a threshold, a detection is classified as positive for PSB
if the difference between normalized pixel intensities exceeds it, otherwise it is classified as negative; here, the threshold may be kept arbitrarily low to ensure a minimum of false negative candidate detections. At last, we add an optional minimum threshold for the relative distance between the base and detected line to exclude false positives as a result of arbitrarily small segments - if it is too small, the detector defaults to a negative classification as a result of a lack of prominent lines.
The approach is summarized in pseudocode in Algorithm 3.\\

\section{Results}
For subsequent tests, we consider two datasets of bias triangles: one simulated dataset with a variety of triangle shapes, obtained by randomizing all parameters per instance run (as per the simulator in prior work \cite{psb1}), and one experimental dataset 
obtained from four silicon FinFET devices with different gate dimensions (a publicly available dataset\cite{psb1}\cite{data}). 
As such, we consider a variety of challenges, in particular, the present FinFET bias triangles often feature prominent excited state lines, but the rest of the shape interior can vanish into the background, while  simulated data features less prominent lines, which on the one hand enables less complex segmentations, but may present challenges for robust line detection. \\
We distinguish between data-dependent and data-independent sets of parameters. In particular, as part of the segmentation algorithm, the choice of thresholding method can be made based on known characteristics of the histogram function of the images in a device-type specific dataset: for the FinFET dataset, we select the mean method due to frequent occurrence of irregular shapes, noise and illumination, and for the 
simulated datasets the triangle method due to the distinct separation between shape and background peaks. 
Further, we set the image interpolation factor to $R = 2$ for the 
simulated dataset, and $R = 4$ for the FinFET dataset - this is to enhance segmentation performance in light of low resolution, where, in particular, the FinFET data includes a larger range of degraded quality samples. The ultimate choice therefore depends on the image quality, however, a value between 2-4 is usually robust.
Further data-independent parameters are universally set as: slope deviation tolerance slope\_tol = 0.4, PSB-threshold int\_tol = 0.05, while the triangle direction is supplied by the experimenter; for brevity, in the present scenario all triangles point in the upward direction.
Here, slope\_tol is selected in view of the fact that the lines detected by the FLD may encompass small segments at high-contrast boundaries which are not perfectly parallel to the base - to limit misleading detections we freely select a value below 1. Normalized intensity values mainly range from 0 to 1, and, accordingly, the PSB-threshold int\_tol is chosen low arbitrarily.

\subsection{\label{sec:level2} Segmentation \& Feature Extraction}
Simulated data was generated together with pixelwise ground-truth annotations for segmented regions of interest based on a physics-inspired simulator \cite{psb1}. Having ground-truth pixelwise labels enables the direct measurement of segmentation performance using metrics that compare pixel-by-pixel the segments extracted by the algorithm with ground-truth, pixel-wise labels.
We assess the segmentation performance on the basis of the Intersection over Union (IoU) and Dice coefficient metrics, respectively defined as
\begin{eqnarray}
Dice = \frac{2|X\cap Y|}{|X|+|Y|}\;
\\
IoU = \frac{X\cap Y}{X \cup Y}
\label{eq:one}
\end{eqnarray}
\[\text{$X\in\mathrm{R}^{N\times M}$: ground truth mask,\enskip $Y\in\mathrm{R}^{N\times M}$: extracted mask}\]
on simulated data with pixelwise ground truth shape masks, exemplified in Table ~\ref{tab:table1} - as an added degree of difficulty, the generated data is further set to include switch-noise, emulating realistic shape interruptions due to charge switch and drift-related movements in gate voltage space. \\
For experimental data stemming from the FinFET device, where pixelwise annotations are not available, we showcase the goodness of segmentation via select visual examples in Figs.~\ref{fig:wide} and ~\ref{fig:wide2}. \\
As a result of the lack of available manual annotations for features such as the base and/or excited state lines, we do not conduct a numerical evaluation, but showcase select examples. Due to the fact that the segmented shape is reduced to a polygon with a minimum number of edges, the majority of feature extraction methods utilizes a succession of simple geometric rules, relying solely on the initial segmentation result, and as such, do not require explicit verification.
\\
We observe in Table \ref{tab:table1} that segmentation as well as classical pixel-classification scores are robust, with an average Dice coefficient of 87.03\% on 500 diverse simulated bias triangles; here, we conduct no interpolation.
In addition, we remark that the rRDP uses the IoU as an inbuilt segmentation score, which ensures that in cases of heavy illumination and/or ambiguous boundaries and shape distortions, the area to be fitted does not deviate significantly from the originally detected area.\\
Further, we discern from the visual examples in Figs.~\ref{fig:wide} and ~\ref{fig:wide2} that the approach reliably segments a variety of different shapes with different image qualities and on this basis, features such as the base line and detuning axes can be automatically extracted. 
\begin{table}
\caption{\label{tab:table1}Average segmentation metrics in \% for the simulated dataset featuring 500 randomly generated cases. All metrics are pixelwise.}
\begin{ruledtabular}
\begin{tabular}{c|cc|ccc}
Data source & IoU & Dice & Accuracy & Precision & Recall \\
 \hline \\
  Simulated & 78.20 & 87.03 & 97.92 & 85.37 & 91.16 \\
\end{tabular}
\end{ruledtabular}

\end{table}

\subsection{\label{sec:level2}PSB-detection}
\begin{figure}

\includegraphics[scale=0.35]{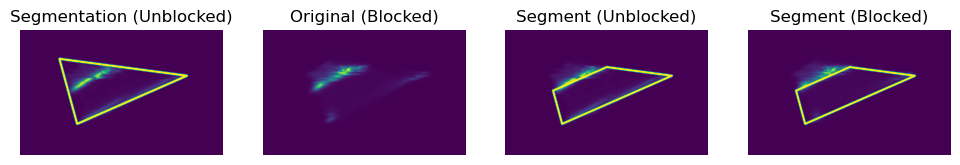}

\includegraphics[scale=0.35]{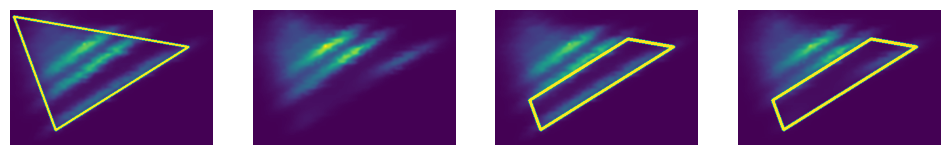}

\includegraphics[scale=0.35]{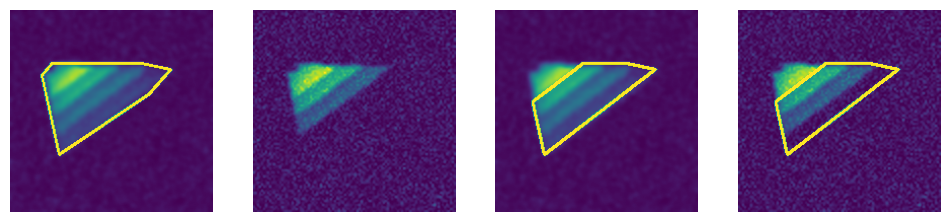}

\includegraphics[scale=0.35]{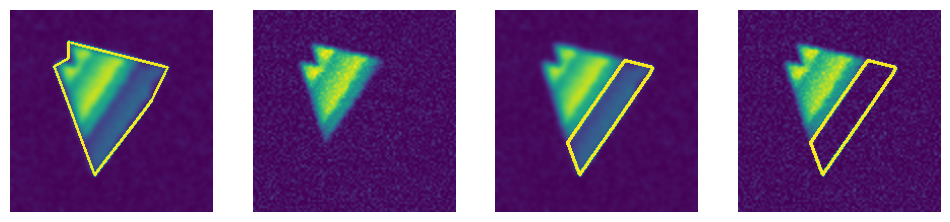}

\caption{\label{fig:psb} Bias triangle pair at different magnetic fields $B$, positively detected as exhibiting PSB with segmented regions of interest displayed, from left to right: in unblocked form with full segmentation at $B\neq0$, in blocked form at $B=0$, respectively in unblocked and blocked form with the segment used to detect PSB. Samples were taken from the FinFET (first \& second row), and simulated datasets (third \& fourth row).}
\end{figure}

We conduct testing of the proposed PSB-detector on both experimental and simulated datasets, featuring a variety of shapes and data quality, and report results in Table ~\ref{tab:table2} in form of accuracy, recall and precision score metrics, as well as the confusion matrix in order to provide a direct overview of individual detections.\\
It becomes evident that the method achieves high classification scores approximately ranging from mid-80\% to high-80\%, with the top score achieved on the simulated dataset. In particular, the FinFET data set, which features a wide range of qualities, as well as the larger simulated dataset, exhibit high classification scores with relatively few false positive and false negative detections, while capturing a wide range of data cases.\\
Furthermore, in Fig.~\ref{fig:psb} we display a selection of bias triangle pairs, respectively measured at different magnetic fields, along with their segmentations, which represent true positive detections of PSB, across different datasets. We observe that the targeted extraction of a segment enclosing the base and excited state accurately captures the differences in intensity due to PSB.

\begin{table}
\caption{\label{tab:table2} PSB Classification Metrics (in \%) for different datasets; we include the confusion matrix $\begin{bmatrix}
 TN & FP\\
 FN & TP\\
 \end{bmatrix}$ with True Negatives (TN), False Positives (FP), False Negatives (FN) and True Positives (TP). The simulated, and FinFET datasets respectively feature 500, and 53 samples.}
\begin{ruledtabular}
\begin{tabular}{c|c|c|c|c|c|c|c}
 Data source & Accuracy & Recall  & Precision & Confusion Matrix 
 \\
 \hline \\
 Simulated &88.2&87.4&89.16& $\begin{bmatrix} 219 & 27\\32& 222\end{bmatrix}$\\
 FinFET &83.02&89.47&70.83& $\begin{bmatrix}27& 7\\ 2& 17\end{bmatrix}$\\
 \end{tabular}
\end{ruledtabular}
\end{table}

\section{Conclusion}
In this work, we have introduced an automatic method to detect, fit and extract geometric as well as physical features from pairs of bias triangles and demonstrated its effectiveness on experimental and simulated datasets. We have built upon and put these properties to concrete use by devising an unsupervised, segmentation-based PSB-detector, which can detect even subtle occurrences of the phenomenon through its localized approach.\\
While the presented framework is efficient and robust for the presented data cases, it still requires further refinement to reliably detect any potential outliers and edge cases, i.e. the occurrence of multiple distinct pairs of bias triangles of different quality in the same stability diagram as well as illumination of varying quality (i.e. current bounds) within the same dataset. While for certain devices and experiment runs, hyperparameters can be reliably determined in advance and fixed, certain device data such as from different FinFET devices, can feature a lot of variability in quality and shapes. \\
Future work may therefore entail the optimization of the decision process between employing rRDP or MET for optimal segmentation. Moreover, present and future work considers the integration of the automatic detection and feature extraction framework as an enabler in an automated tuning pipeline.
\begin{acknowledgments}
We gratefully acknowledge the support of Prof.\ Natalia Ares, University of Oxford.
In addition, this work is supported by the Innovate UK, grant number 1004359. J.S. acknowledges financial support from the EPSRC, grant Number R72976/CN001.
\end{acknowledgments}

\section*{Code Availability Statement}
The code will be made available upon final publication.

%\appendix

%\section{Appendixes}

\nocite{*}
\bibliography{main_paper}% Produces the bibliography via BibTeX.

\end{document}